\documentclass[prl,aps,twocolumn,groupedaddress,floats,showpacs]{revtex4}
\usepackage{graphicx}
\usepackage{dcolumn}
\usepackage{bm}
\def\he4{$^4$He}
\def\Am3{\AA$^{-3}$}
\def\beq{\begin{equation}}
\def\eeq{\end{equation}}
\begin{document}

\title{Superglass Phase of \he4}
\author{Massimo Boninsegni$^{1}$, Nikolay Prokof'ev$^{2,3,4}$, and
Boris Svistunov$^{2,3}$}
\affiliation{ ${^1}$Department of Physics, University of Alberta,
Edmonton, Alberta T6G 2J1\\
${^2}$Department of Physics, University of
Massachusetts, Amherst, MA 01003 \\
${^3}$Russian Research Center ``Kurchatov Institute'', 123182
Moscow \\
${^4}$Department of Physics, Cornell University,
Ithaca, NY 14850}
\date{\today}
\begin{abstract}
We study different solid phases of \he4, by means of Path Integral
Monte Carlo simulations based on a recently developed {\it worm}
algorithm. Our study  includes simulations that start off from a
high-$T$ gas phase, which is then ``quenched" down to
$T$=0.2 K. The low-$T$ properties of the system crucially depend
on the initial state. While an ideal {\it hcp} crystal is a
clear-cut insulator, the disordered system freezes into a  {\it
superglass}, i.e., a metastable amorphous solid featuring
off-diagonal long-range order and superfluidity.
\end{abstract}

 \maketitle

The remarkable observation by Kim and Chan of a non-classical moment
of inertia in solid $^4$He \cite{KC} has generated a new wave of
interest in the  possible superfluid phase of a solid.
Supersolidity of $^4$He is still controversial, both at the
experimental and theoretical levels. Two of us have recently
proven that, irrespective of its microscopic structure,
any supersolid crystal should contain gapless vacancies
and/or interstitials  \cite{theorem}. In other words, any
continuous-space supersolid is generically {\it incommensurate}
(i.e., the number of atoms per unit cell is not an integer)
and {\it squeezeable}, i.e., by applying pressure it should be
possible to squeeze matter from a container with supersolid
into a buffer volume containing the same supersolid.
However,  this very experiment has yielded a negative
result for solid \he4 \cite{Bi}.

A wealth of numerical studies clearly indicate that
$^4$He is a commensurate (thus insulating) crystal.
The finite activation energy of a vacancy computed numerically is
large, $\sim$ 15 K, and claimed consistent with the experimental
observations \cite{Galli}. The activation energy of an
interstitial, $\sim 50\,$K \cite{CB}, is significantly
larger than that of a vacancy. A simulation study of
exchanges in an ideal {\it hcp} crystal \cite{CB}, yielded
indirect evidence that the system is not superfluid.
In sharp contrast, the variational ($T$=0)
calculation of Ref. \onlinecite{Galli_2} claims a finite
condensate fraction in the {\it commensurate} $^4$He crystal.
Thus, additional investigation is  warranted.

The experiment of Kim and Chan itself has revealed a number of
facts pointing to a strongly inhomogeneous scenario of
superfluidity, chiefly the contaminating effect of a small
concentration of $^3$He, and non-XY behavior of the superfluid
density at the critical temperature.
The need of exploring inhomogeneous (metastable) scenarios of
supersolidity, dictated both by theory and experiments, has
already resulted in some relevant theoretical developments,
revealing superfluid interfaces in a lattice solid
\cite{interfaces} and a superfluid layer at the boundary between
the $^4$He crystal and a disordered substrate \cite{KhC}.

The  numerical observation of a
metastable disordered supersolid, (a {\it superglass} phase of
$^4$He) is reported in the present Letter. To be specific in the
definition, by  {\it glass} we mean a spatially disordered
(metastable) phase,  indistinguishable from a solid \cite{def} on
a time scale much shorter  than the typical relaxation time, $t_{\rm
rel}$, which in turn should be dramatically longer than the
inverse Debye frequency, $\omega_D^{-1}$. Superglass is the term
that we use for such a phase, if it also displays superfluidity.
Note that our definition of glass does not address the behavior of 
the system at time scales $t \gtrsim t_{\rm rel}$, whereupon it may 
undergo structural relaxation into the polycrystalline
sample, or simply behave as a very viscous liquid.

In Ref.~\cite{balibar}, the idea of glassification of
overpressurized liquid \he4 was put forward, in order to explain a
striking experimental outcome, i.e., the absence of bulk solid
nucleation under fast (about $1~\mu s$) acoustic wave compression
pulses, up to pressures as high as $\sim 160\,$bar. The authors
conjectured that the glassy phase is {\it normal} (though the
experiment was done at $T=0.05$~K and the adiabatic heating was
estimated to be below $0.1$~K);  the absence of superflow towards
the nucleation center would dramatically suppress the rate of
growth of the crystal. Conceptually, the finding of the present
Letter is different, but we believe relevant to the interpretation
of the experiment of Ref.~\cite{balibar}. Jamming of structural
relaxation does not  {\it per se} exclude superfluidity.
Crystallization is suppressed by the mere fact that the normal
component forms a glassy solid, implying that further evolution
towards a lower-energy polycrystal structure necessarily involves
a chain of exponentially rare quantum-tunnelling or
thermoactivation events, rather than a rapid growth of the
supercritical nucleus. Indeed, the boundary between the
perfect-crystal nucleus and the superglass, is a solid-solid
interface which realizes a pronounced local energy minimum. Its
evolution should therefore imply either quantum tunnelling (in the $T
\to 0$ limit), or thermoactivation.

Our study is based on accurate Path Integral Monte Carlo (PIMC)
simulations of condensed $^4$He, making use of a recently developed worm
algorithm \cite{worm2005}. This method allows for efficient
sampling and accurate determination of the single-particle Green
function and superfluid density, for systems comprising a
relatively large number $N$ of particles (of the order of several
thousand). Specifically, we address the following two issues: (i)
Is it possible to obtain definitive first-principle theoretical
evidence that an ideal {\it hcp} \he4 crystal is an insulator ?
(ii) What happens to a sample of liquid \he4 quenched through the
first-order liquid-solid phase transition?

We consider a system of $N$ \he4 atoms ($N$=216 and 800), at a
temperature 0.2 K $\le T \le$ 1 K,  and at the two densities
$n$=0.0292 (0.0359) \Am3,  corresponding to an ideal {\it hcp}
\he4 crystal at a pressure of approximately 32 (155) bars
\cite{Driessen}. The sample cell geometry with periodic boundary
conditions is designed to fit an ideal {\it hcp} crystal. We use
the standard microscopic model of
 \he4, based on the Aziz pair potential \cite{aziz}.

\begin{figure}[tbp]
\centerline{\includegraphics [bb=40 90 590 730, angle=-90,
scale=0.34]{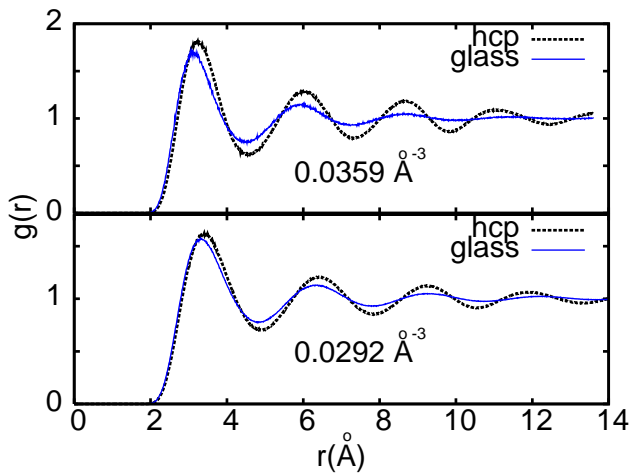}} \vspace*{-0.4cm} \caption{(Color online).
Pair correlation function of the ideal \he4 {\it hcp} solid and
superglass at the near-melting ($n$=0.0292 \Am3, lower panel) and
higher ($n$=0.0359 \Am3, upper panel) densities.} \label{f1}
\end{figure}

\begin{figure}[tbp]
\centerline{\includegraphics[angle=-90, scale=0.32] {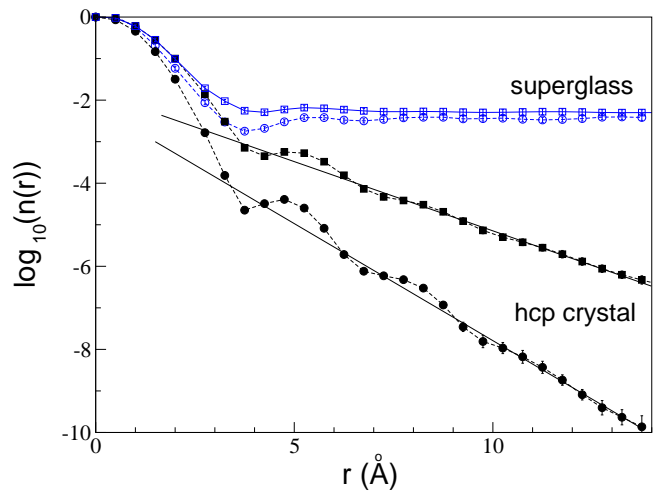}}
\vspace*{-0.4cm} \caption{(Color online). Single-particle density
matrix $n(r)$ for ideal \he4 {\it hcp} crystal (filled symbols)
and superglass (open symbols) at the near-melting density
$n$=0.0292 \Am3 (squares) and high-density $n$=0.0359 \Am3
(circles). Solid lines through filled symbols represent
exponential decay. } \label{f2}
\end{figure}

In Figs. \ref{f1} and \ref{f2}, we show data for the pair
correlation function $g(r)$ and the single-particle density
matrix $n(r)$. For both the near-melting density of $n$=0.0292 \Am3 and
the higher density of  $n$=0.0359 \Am3 we study {\it two} samples,
differing in one respect only, namely their initial
configurations before equilibration.

The single-particle density matrix is defined as 
$n({\bf r},{\bf r}')\; =\;  \langle\, \hat{\psi}^{\dag}({\bf
r})\: \hat{\psi}({\bf r'}) \, \rangle$ 
where
$\hat{\psi}({\bf r})$ is the particle annihilation operator, $\hat
{\rho}({\bf r})= \hat{\psi}^\dagger({\bf r}) \hat{\psi}({\bf r})$ is
the local \he4 density operator, and $\langle ...\rangle$ stands
for thermal average. It is customary to display the spherically averaged
function \beq n(r) = \frac{1}{4\pi V} \int d\Omega\ \int d^3r'\
n({\bf r'}, {\bf r'}+{\bf r}) \eeq where $V$ is the volume of
the system. This is the quantity shown in Fig. \ref{f2}.

When the simulation is started from an initial configuration
corresponding to an ideal {\it hcp} crystal, we consistently find
an exponential decay of $n(r)$ at large distances, with
short-range oscillations due to coordination-sphere effects.
We observe no change in the results between the temperatures
of 0.2 and 1 K, to indicate that those shown in Fig. {\ref {f2}}
are essentially ground state estimates.
This result provides a robust confirmation that an ideal {\it hcp}
crystal is {\it not} a Bose condensate (superfluid).
This conclusion is consistent with the theoretical
expectation that a crystal with finite activation energies for
vacancies and interstitials will not display superfluidity
\cite{theorem}, and is in agreement with arguments based on the
statistics of exchange cycles observed  in the same system
\cite{CB}. The results and
conclusions of Ref.~\cite{Galli_2} appear therefore to be
erroneous, possibly artifacts
of the variational approach.

At present, there is no clear understanding of what crystalline
defects dominate in the experimental samples of
Ref.~\onlinecite{KC}. It is not known whether individual
dislocations, dislocation sheets and networks, or grain boundaries
in bulk $^4$He may underlie  the experimentally observed
superfluid response, though model simulations of domain walls in
quantum solids hint at such possibilities \cite{interfaces}.
But regardless of their nature, in the absence of crystalline
defects no theoretical interpretation seems viable of the
experiments reported in
Ref.~\onlinecite{KC}, in terms of superfluid response.

In order to investigate scenarios of broken translation
invariance not involving a perfect crystal 
(though perhaps not
directly related to the experiment of Ref. \onlinecite{KC}), 
we designed a
simulation protocol aimed at mimicking a ``quenching" experiment
(namely one in which liquid Helium is suddenly, rapidly cooled)
obviously making allowance for the important differences between
the imaginary-time PIMC dynamics and the real-time dynamics
of actual physical systems (see discussion below).
Starting from an initial configuration characteristic of a high-$T$ gas phase,
we ``quench" the system down to the temperature $T=0.2$~K by
$>10^{4}$ PIMC sweeps. One ``sweep" is defined as
the number of accepted updates sufficient to sample the entire
Path Integral configuration. We then run the simulation
long enough to achieve stability of statistical averages for
structural properties and for the single-particle
density matrix.

The phase that emerges from disorder resembles the {\it hcp}
crystal only at short interatomic  distances (of about
two-three coordination spheres), with no diagonal long-range order
(see Fig. \ref{f1}). Moreover, this phase has a well-developed
off-diagonal long-range  order, with a condensate fraction
$n_\circ =0.5~\%$ (see Fig.~\ref{f2}), and a surprisingly large
superfluid fraction $\rho_s=0.6(1)$ at $n$=0.0292 \Am3 and
$\rho_s=0.07(2)$ at $n=0.0359$ \Am3. Though the superfluid
fraction is strongly suppressed with pressure the condensate
fraction is reduced by merely $50\%$.

The important observation  is that \he4 can remain in the
metastable superfluid state at solid-state densities, even at
fairly high pressure. This observation is consistent with a previous
study \cite{moroni04}, predicting  overpressurized {\it liquid} 
\he4 to remain superfluid (at $T$=0)  to arbitrarily high density. 
The nature of the superfluid phase will generally 
depend on pressure, temperature, and the experimental time scale.
For example, one may expect that the lower density 
finite-temperature phase should be just a superfluid,  but with a rather 
viscous normal component. On the other hand, with
increasing density such a normal component may evolve into a glass, 
with a diverging (i.e., unobservably large) viscosity.

In order to study whether and how the system breaks translation symmetry, 
we calculate the condensate wave function $\phi({\bf r})$. 
The worm algorithm offers direct access
to the one-body density matrix which in the presence 
of off-diagonal
long-range order factorizes  at large separation
$|{\bf r}-{\bf r'}|$
\[
\langle\,  \hat{\psi}^{\dag}({\bf r})\: \hat{\psi}({\bf r'})\,
\rangle \; \to\;  \phi ({\bf r}) \: \phi ({\bf r'}) \; .
\]

\begin{figure}[tbp]
\centerline{\includegraphics [bb=68 75 592 728,
width=1.98\columnwidth  ]{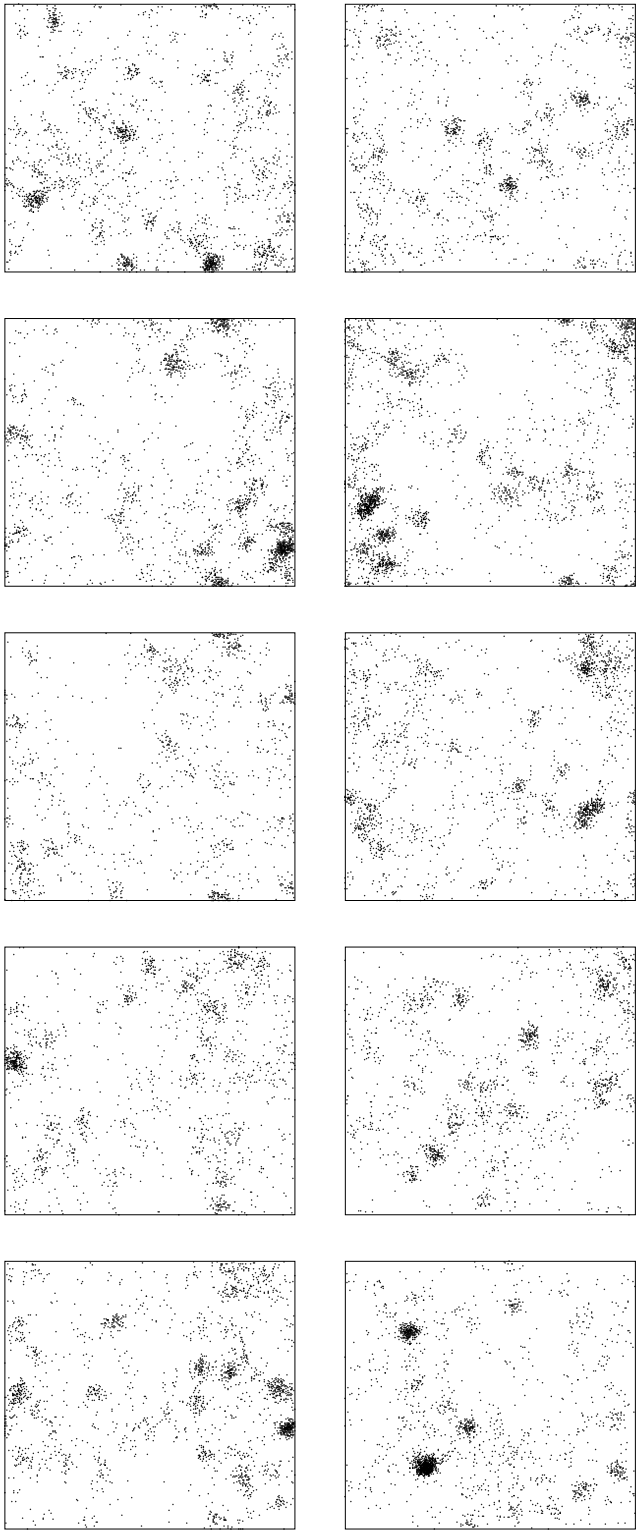}} \caption{Condensate wavefunction
at $n$=0.0359 \Am3 
(represented by the density of points) and $T$=0.2 K, obtained by making ten
slices of the system along the $\hat{z}$-direction and projecting
them on the $xy$-plane (slices are ordered from left bottom to
left top and then from right bottom to right top).} \label{f5}
\end{figure}

In Fig.~\ref{f5}, we show two-dimensional $xy$-maps of the condensate
wave function $\phi(x,y,z)$ at $n=0.0359$ \Am3, for ten (equally spaced) slices
along the $z$-axis. The data shown in Fig.~\ref{f5} represent
long simulation-time averages, not instantaneous snapshots. Aside from
the obvious observation that the system has manifestly broken
translational invariance, the results suggest no obvious
interpretation of the disordered pattern
for $\phi({\bf r})$ in terms of dislocations or grain
boundaries \cite{note3}. Correlations barely extend over two
slices in Fig.~\ref{f5}. The conclusion that we draw from
Figs.~\ref{f1}-\ref{f5}, is that $^4$He forms a  superglass.

Naturally, the results shown in Fig.~\ref{f5} are
influenced by a particular gas-like initial condition;
another initial condition would produce a
different result for $\phi({\bf r})$. Nonetheless, the fact that
superfluidity and off-diagonal long-range order appear for just
{\it one} such random initial condition, strongly suggests that
these are genuine physical properties of the metastable phase.

Despite the above-mentioned fact that the MC dynamics is
quite different from the real-time dynamics of helium (in
simulations heat dissipates locally) it is possible to make
semiquantitative arguments with regards to the stability of the
superglass phase. Its stability on timescales several orders of
magnitude longer than $\omega_D^{-1} \sim 3\time 10^{-13}$ s,  
is guaranteed by observing no changes in the superglass properties over $10^4$ MC sweeps. 
[The most appropriate physical interpretation of one sweep for the conventional 
PIMC scheme, is the time scale corresponding to the zero-point
motion of atoms, whereby all particles have a chance to sample 
their optimal positions locally.] 
The extra advantage of using the worm algorithm
is that pair-wise exchange, or tunneling of two particles,
is sampled at the same rate as zero-point vibrations,
while the rate $J$ of exchange processes in the solid state
of \he4 is about five orders of magnitude slower 
(e.g., measured values of tunneling for $^3$He in  $^4$He are of order 
$J \sim 10~\mu s^{-1}$ \cite{Richards}). 
It seems then plausible to assume that the metastability
of the superglass phase extends up to $10^4
J^{-1} \sim 1~{\rm ms}$. If multiparticle tunneling (at
low-temperature) events are required to reach the genuine
equilibrium, then the actual degree of metastability has no
obvious upper limit, and can easily exceed the
experimental timescale.

Summarizing, we have provided theoretical evidence that $^4$He
features a new metastable phase, a superfluid glass. This
observation naturally suggests that other, more ``regular" types
of solid disorder, such as grain boundaries and dislocations, may also
possess superfluid properties. We foresee further
theoretical studies in the following directions:
\begin{enumerate}
\item {Determination of the ``phase diagram" of the glassy phase,
i.e., of its domain of metastability (e.g.,
in the $(n,T)$ plane)  and the line separating superglass from
normal glass. The other important issue is to quantify the crossover line separating this novel, superglass phase from the superfluid: the superglass phase is characterized by the 
low-temperature plateau $\rho_s(T)\to\rho_\circ < 1$, as in a ``dirty" conventional superfluid.}
\item{ Explore alternative possibilities of
metastable supersolids: (i) A regular crystal doped with vacancies
(or even interstitials). (ii) Superfluid grain boundaries and/or
dislocations. (iii) The yet elusive superfluid phase of condensed
{\it para}-hydrogen.}
\end{enumerate}

The superglass phase is not {\it directly} relevant to the interpretation of Kim and Chan's experiment, since the MC temperature quench is much more rapid then in the experiment, leading to more disordered samples and much larger 
superfluid fraction.  
However, a related scenario may be appropriate, namely that of a generalized superfluid-grain-boundary \cite{interfaces}, 
which may may include a foam-shaped superglass network interpenetrating the polycrystal.

We acknowledge valuable discussions with E. Mueller, V. Essler, J.
Reppy, B. Guyer, B. Mullin, B. Hallock, D. Candela, S. Girvin, and
A. Leggett. This work was supported by the National Aero and Space
Administration grant NAG3-2870, the National Science Foundation
under Grants Nos. PHY-0426881 and PHY-0456261, by the Sloan
Foundation, and by the Natural Science and Engineering Research
Council of Canada under grant G121210893.

{\it Note added.} In an independent study \cite{clark}, Clark
and Ceperley calculated the single-particle density matrix of the
ideal {\it hcp} \he4 crystal at the melting density
$n=0.0287$\Am3. The data of Ref.~\cite{clark} are consistent
with ours for $n=0.0292$\Am3.


\begin{thebibliography}{99}
\bibitem{def} By definition, a solid is a state with broken
translation symmetry. An immediate implication of broken
translation symmetry is shear rigidity.
\bibitem{note3} In any case, small scale and the amount
of disorder would make any such interpretation rather ambiguous.
\end{thebibliography}
\end{document}